# Molecular Dynamics Study of the Mechanical Behavior of Few Layer Graphene

*Young In Jhon[1] and Myung S. Jhon[1,2]\**

[1]School of Advanced Materials Science and Engineering, Sungkyunkwan University, Suwon 440-746, Korea, [2]Department of Chemical Engineering and Data Storage Systems Center, Carnegie Mellon University, Pittsburgh, PA 15213, USA

**Abstract**

Atomistic simulation was performed to study the mechanical properties of few layer graphene (FLG) in conjunction with monlayer graphene (MLG) under uniaxial elongation by systematically increasing the layer number from one to six. We found that the ultimate tensile strength and strain increased in these FLGs for both zigzag and armchair-directional elongations when compared with the results of MLG. We also found that the largest increments were obtained in bi- or tri-layer graphene for all the FLG systems we studied. Using atomic stress distribution analysis, it is observed that the width of the distribution became narrower, thus the maximum stress decreased in FLG compared to MLG at respective stages of identical tensile stress. It indicates that locally-driven highly elevated atomic stress of FLG has been effectively relaxed to the atoms in other layers through cooperative interlayer interaction. This effect explains the reason for synergetic mechanical

---

\* *Corresponding author.* Fax: + 412-268-7139. Email address: mj3a@andrew.cmu.edu (M.S. Jhon)

strengthening of FLG since tensile failure is critically influenced by maximum atomic stress. Furthermore, the Young's moduli were slightly smaller for all FLGs compared to MLG.

## 1. Introduction

A mono atomic layer of graphite known as graphene has been attracting great interest due to its exceptional properties and broad spectrum of applications. As a novel two-dimensional material composed of hexagonally $sp^2$-bonded carbon atoms, it exhibits ultrahigh electron mobility which is 10 times more than that of silicon wafers [1,2], superior thermal conductivity [3,4], and excellent mechanical strength including exceptional stretchability [5-7]. While most attention has been paid to monolayer graphene (MLG) until now [8-10], it was recently found that stacks with a finite number of MLG sheets known as few layer graphene (FLG) exhibited many intriguing features which are not obtained from its MLG form [11-14]. Especially, a large number of studies have been performed on band gap opening induced by electric field in bilayer graphene [15-19] and subsequently, extensive research is being done to commercialize this phenomenon [20-23]. FLG can be synthesized either by exfoliating graphite [24] or by epitaxially growing graphene layers [25]. However, as the mechanical properties of FLG cannot be obtained easily due to difficulties in experiments for delicate manipulation & precise measurement, detailed understanding of its mechanical characteristics has been still left as an unsolved problem despite its critical importance in modern industrial applications.

To tackle this unresolved issue, we performed molecular dynamics (MD) simulations to investigate the mechanical behaviors of FLG sheets under zigzag (ZZ) and armchair (AC)-directional elongations by systematically varying the layer numbers from two to six. In this study, these systems were denoted as *dir-n*L where *dir* denotes the direction of the elongation, namely, ZZ or AC and *n*L implies the layer number. The atomic stress distribution analysis was also performed to understand the physical interpretation of our results.

## 2. Computational Methods

MD simulations were performed using the software package LAMMPS [26] with the adaptive intermolecular reactive empirical bond order (AIREBO) potential [27] and a time step of 1.0 fs. Rhombohedral stacking, namely, the stacking order of *ABCABC*... was adopted to construct the structure of FLG. The periodic dimensions of the simulation system and atomic coordinates were first optimized using a gradient-based minimization method with tolerance criteria of $10^{-8}$ eV/Å for force and/or $10^{-8}$ eV for energy. Based on the simulation cell size obtained from the above calculation, which was 50 Å $\times$ 50 Å $\times$ 30 Å, an NVT simulation was performed consecutively for $3\times10^5$ steps. Then, the system was elongated in the ZZ or AC-direction. To make a refined structural variation during elongation, the strain rate was set to 0.1 ns$^{-1}$ and 2 000 steps were taken at each deformation point, where accordingly, a 0.02% strain was applied to the system between two consecutive points. Non-equilibrium molecular dynamics (NEMD) simulations of a continuously strained system were carried out with SLLOD equations of motion coupled to a Nose/Hoover thermostat [28]. For the AIREBO potential, the cut-off radius parameter was set to be 2.0 Å to avoid spuriously high bond forces and unphysical results near the fracture region. This value has tacitly been used in most of the precedent research in studying the mechanical properties of graphene [29,30].

The atomic stress tensor of individual carbon atoms in FLG was calculated by:

$$\sigma_{ij}^{\alpha} = \frac{1}{\Omega_{\alpha}} \left( \frac{1}{2} m_{\alpha} v_i^{\alpha} v_j^{\alpha} + \sum_{\beta=1,n} r_{\alpha\beta}^{j} f_{\alpha\beta}^{i} \right),$$

where, $i$ and $j$ are components in Cartesian coordinates; $\alpha$ and $\beta$ are the atomic indices; $m_{\alpha}$ and $v^{\alpha}$ are the mass and velocity of $\alpha^{th}$ atom, respectively; $r_{\alpha\beta}$ is the distance between $\alpha$ and $\beta$ atoms, and $\Omega_{\alpha}$ is the volume of $\alpha^{th}$ atom.

After $\sigma_{ij}^{\alpha}$ is obtained, the tensile stress of the system was computed by averaging over all the carbon atoms.

## 3. Results and discussion

The stress-strain curves of MLG and FLGs were obtained under either ZZ or AC-directional tensile load by varying the layer number between 1~6, as shown in Figs. 1 and 2. We found that the ultimate strength and strain of FLG are always greater than those of MLG, regardless of the layer number, implying that FLG exhibits interlayer-based synergetic strengthening. The largest increments in the ultimate strength and strain of FLGs were found in either bi- or tri-layer graphene sheet rather than thicker FLGs for both ZZ and AC-directional elongations as shown in Fig. 3. Compared to MLG, they increased by 2.98 GPa and 0.012 for ZZ-elongation and 4.56 GPa and 0.012 for AC-elongation, respectively. It is noted that this improvement in our result is contrary to the findings for differently structured bilayer graphene where the presence of sp$^3$ bonding between layers was assumed with bond length of 1.54 Å while deteriorating the tensile strength as a consequence [31]. In all of the FLG systems we investigated in this study, the distance between layers is maintained at a value around 3.35 Å, where only non-bonding interlayer interactions exist. These systems closely resemble realistic structures obtained from synthesis process. To elucidate the physical background on this strengthening effect of FLG, we analyzed the histogram of atomic stress distributions for ZZ-1L, ZZ-2L, and ZZ-3L systems at the stages corresponding to the identical tensile stress of 80 GPa as shown in Figs. 4 (a)-(c). We found that the width of atomic stress distribution became narrower and consequently, the maximum atomic stress decreased in ZZ-2L compared to ZZ-1L. Here, it should be noticed that, despite the difference in atomic stresses, the tensile stresses of ZZ-1L and ZZ-2L were identical. This suggests that, compared to ZZ-1L, the atomic stress distribution of ZZ-2L was built by the rise in atomic population at the middle range of the atomic stresses and the decrease at the terminal ends, *i.e.,* around maximum and minimum values of atomic stresses. It further indicates that the locally-driven highly elevated atomic stress has been effectively distributed to other atoms in ZZ-2L than ZZ-1L and accordingly, through this relaxation process, ZZ-2L can accommodate larger tensile strength since tensile failure is critically influenced by the maximum magnitude of atomic stresses. A further study on AC-1L, AC-2L, and AC-3L systems validated this hypothesis as shown in Figs. 4 (d)-(f). The ultimate tensile strength was plotted as a function of standard deviation (SD) of the atomic stresses for those systems and it exhibited a positive proportional relationship as shown in Fig. 5.

Furthermore, the spatial atomic stress distributions of AC-1L and AC-2L were represented in Fig. 6. The maximum and minimum atomic stresses were 115.2 GPa & 34.0 GPa for AC-1L, and 107.6 GPa & 50.4 GPa for AC-2L, respectively. However, for comparison, the maximum and minimum values of AC-1L were taken as the two limits of color range for both AC-1L and AC-2L in common. The result showed a few atoms in dark red and blue for AC-1L but they mellowed for AC-2L. It indicates that the effective stress relaxation occurred in FLG by sharing the stress with atoms in other layers.

In addition, Young's moduli of the FLGs were also obtained as shown in Table 1. Although there were some fluctuations with increasing layer number, they are always smaller than that of MLG for all FLGs. A similar trend in the decrease of Young's moduli was observed in previous work on mechanical properties of grafold for a range of folding numbers, which is an architecture of folded graphene nanoribbon [32].

## 4. Conclusion

We studied the mechanical behavior of FLG under both ZZ and AC-uniaxial elongation using MD simulations by varying the layer number between 2~6. It was observed that FLG exhibited stronger tensile strength and higher ultimate strain for all layer numbers compared to MLG. The maximum values of ultimate tensile strength and strain were obtained in bi- or tri-layer graphene for all the FLG systems investigated here in both ZZ and AC-directional elongations. Utilizing the atomic stress distribution analysis, we also found that the highly concentrated stress has been effectively relaxed to a number of atoms in FLG and consequently, the magnitude of maximum atomic stress decreases compared to MLG, presumably through the synergetic interlayer interaction. Based on these findings, we were able to explain a mechanism of how FLG possess larger tensile strength than MLG since tensile failure is critically affected by maximum atomic stress.

We believe that our findings will contribute greatly in fundamental understanding on mechanical characteristics of FLG and future design of FLG-based devices.

## Acknowledgement

This work was supported by the World Class University program of KOSEF. (Grant No. R32-2008-000-10124-0).

**Figure captions**

Figure 1. The stress-strain curves of (a) 1~3 layer (denoted as ZZ-1L, ZZ-2L, and ZZ-3L, respectively) and (b) 4~6 layer (denoted as ZZ-4L, ZZ-5L, and ZZ-6L, respectively) graphene sheets under zigzag-directional elongations.

Figure 2. The stress-strain curves of (a) 1~3 layer (denoted as AC-1L, AC-2L, and AC-3L, respectively) and (b) 4~6 layer (denoted as AC-4L, AC-5L, and AC-6L, respectively) graphene sheets under armchair-directional elongations.

Figure 3. The enlarged stress-strain curves around tensile failure for various FLGs and MLG under (a) zigzag-directional elongations and (b) armchair-directional elongations.

Figure 4. The atomic stress distributions of (a) ZZ-1L, (b) ZZ-2L, (c) ZZ-3L, (d) AC-1L, (e) AC-2L, and (f) AC-3L systems at the stages corresponding to tensile stress of 80 GPa. For 2L and 3L systems, the frequency is reduced by dividing the original frequency by one and two, respectively, for the comparison. SD denotes standard deviation.

Figure 5. The relationship between SD of atomic stresses and tensile strength for ZZ-$n$L and AC-$n$L systems ($n$=1~3).

Figure. 6. The spatial atomic stress distributions of (a) ZZ-1L, (b) an upper layer of ZZ-2L, and (c) a lower layer of ZZ-2L.

Table 1. Young's moduli for MLG and FLGs under ZZ and AC-directional elongations. All units are in TPa.

|    | 1L    | 2L    | 3L    | 4L    | 5L    | 6L    |
|----|-------|-------|-------|-------|-------|-------|
| ZZ | 1.263 | 0.895 | 1.010 | 0.980 | 1.052 | 0.988 |
| AC | 1.097 | 0.989 | 1.001 | 0.915 | 0.949 | 1.016 |

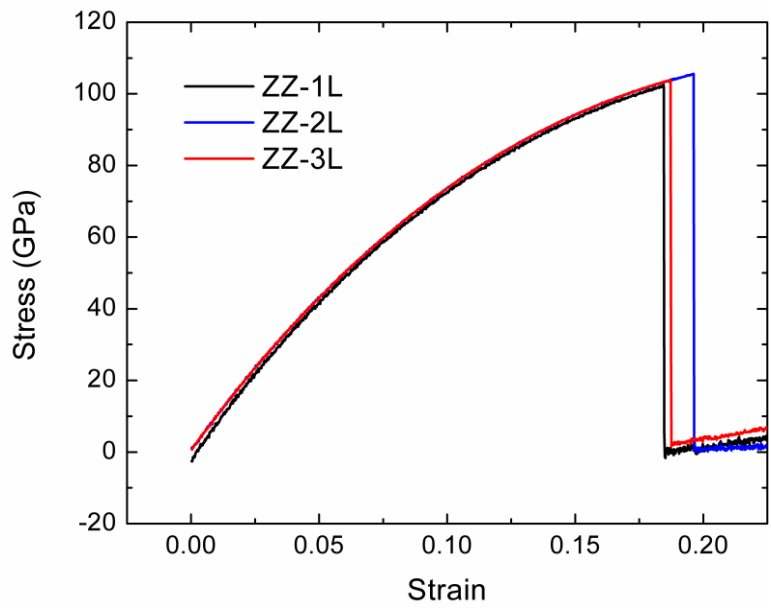

(a)

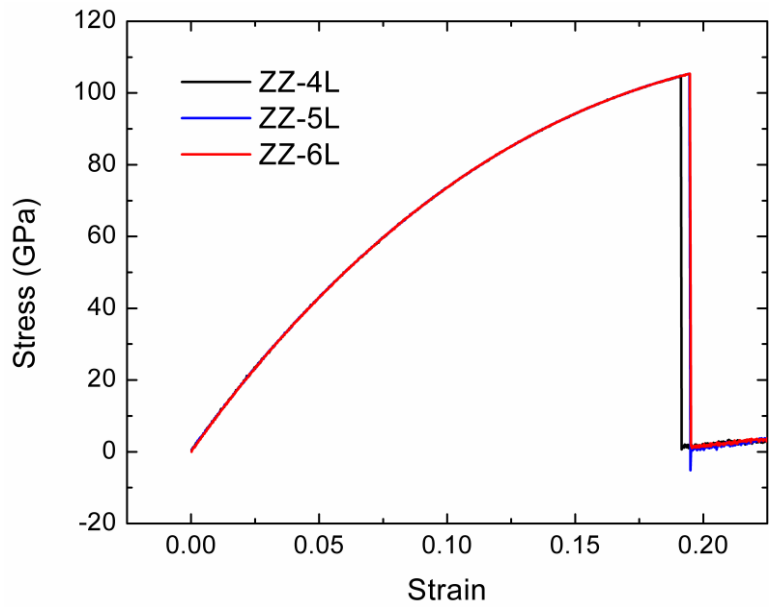

(b)

**Figure 1.** The stress-strain curves of (a) 1~3 layer (denoted as ZZ-1L, ZZ-2L, and ZZ-3L, respectively) and (b) 4~6 layer (denoted as ZZ-4L, ZZ-5L, and ZZ-6L, respectively) graphene sheets under zigzag-directional elongations.

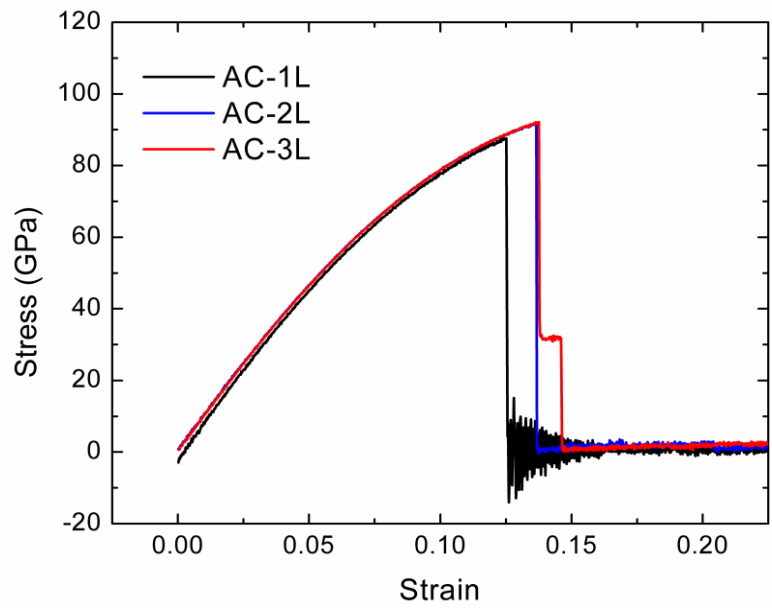

(a)

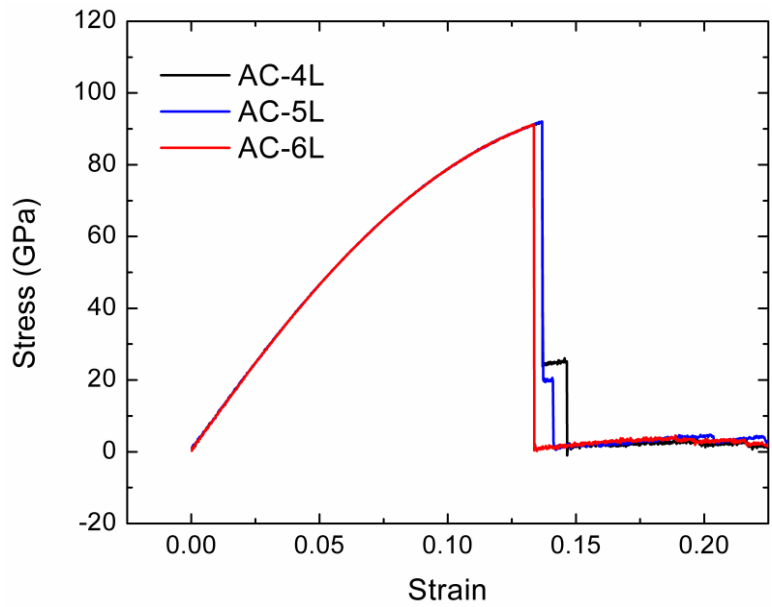

(b)

**Figure 2.** The stress-strain curves of (a) 1~3 layer (denoted as AC-1L, AC-2L, and AC-3L respectively) and (b) 4~6 layer (denoted as AC-4L, AC-5L, and AC-6L, respectively) graphene sheets under armchair-directional elongations.

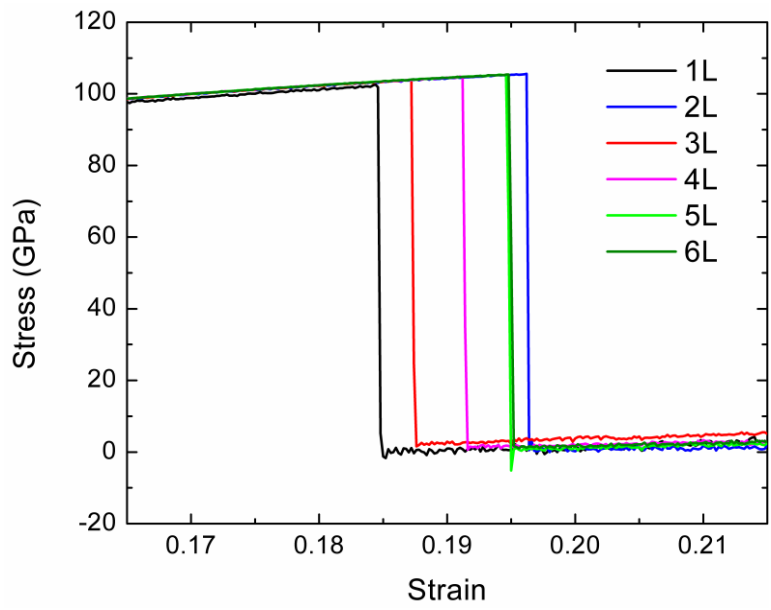

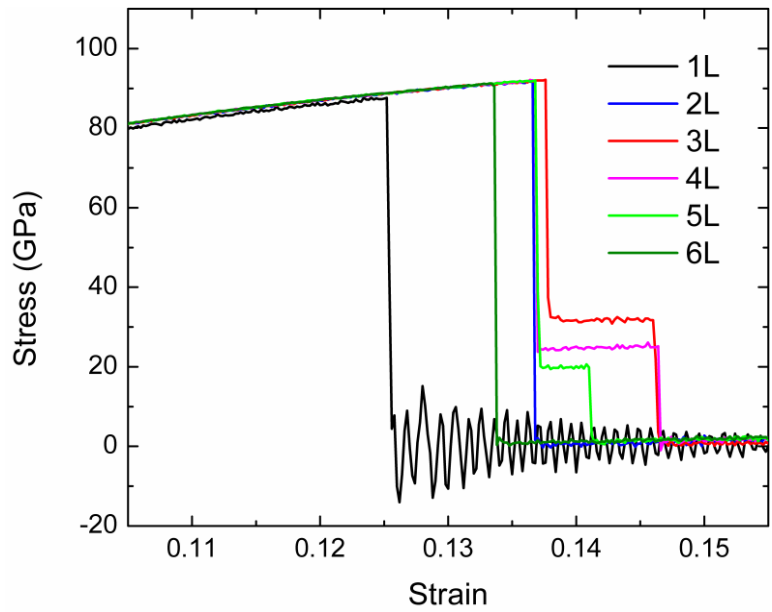

**Figure 3.** The enlarged stress-strain curves around tensile failure for various FLGs and MLG under (a) zigzag-directional elongations and (b) armchair-directional elongations.

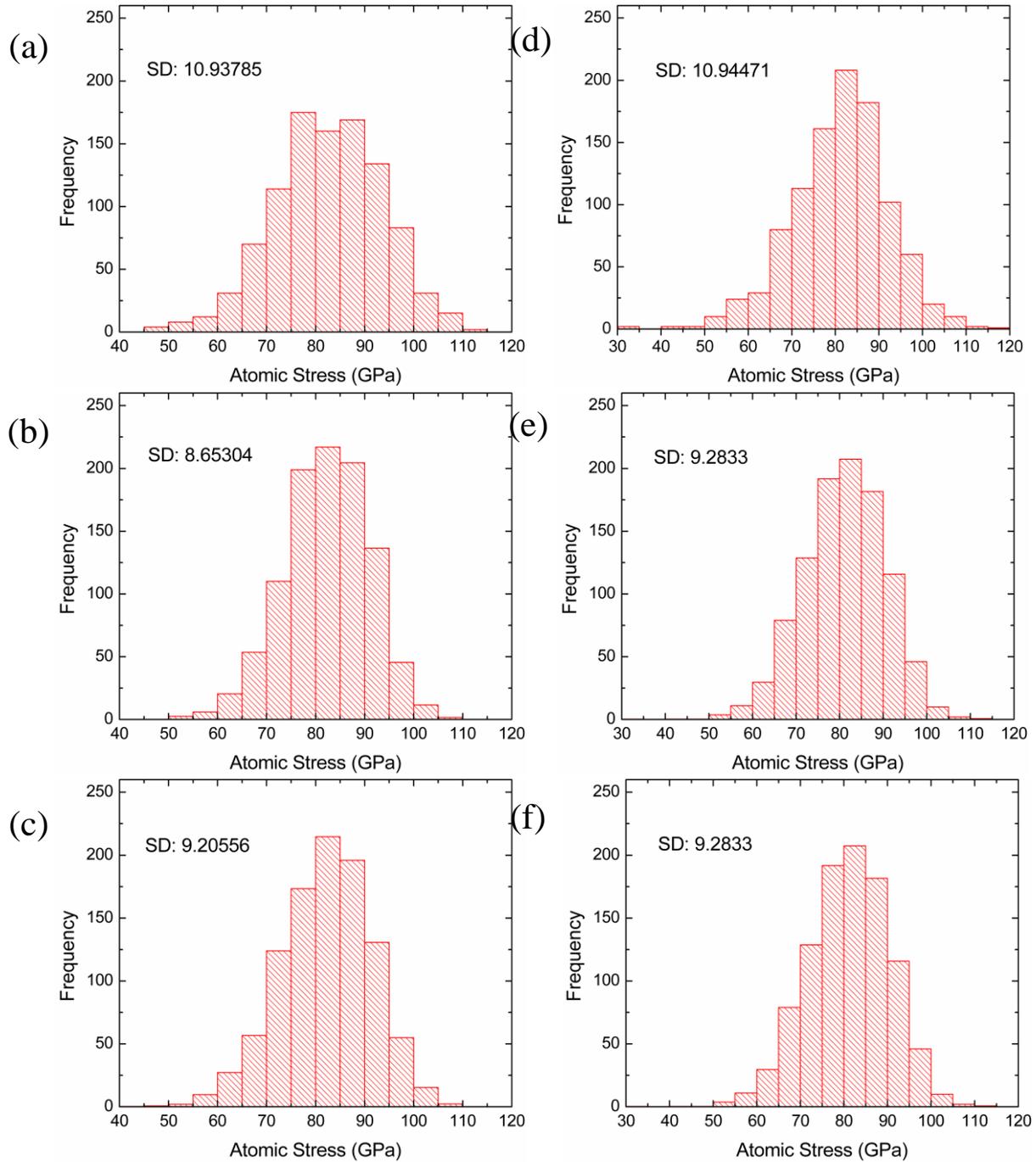

**Figure 4.** The atomic stress distributions of (a) ZZ-1L, (b) ZZ-2L, (c) ZZ-3L, (d) AC-1L, (e) AC-2L, and (f) AC-3L systems at the stages corresponding to tensile stress of 80 GPa. For 2L and 3L systems, the frequency is reduced by dividing the original frequency by one and two, respectively, for the comparison. SD denotes standard deviation.

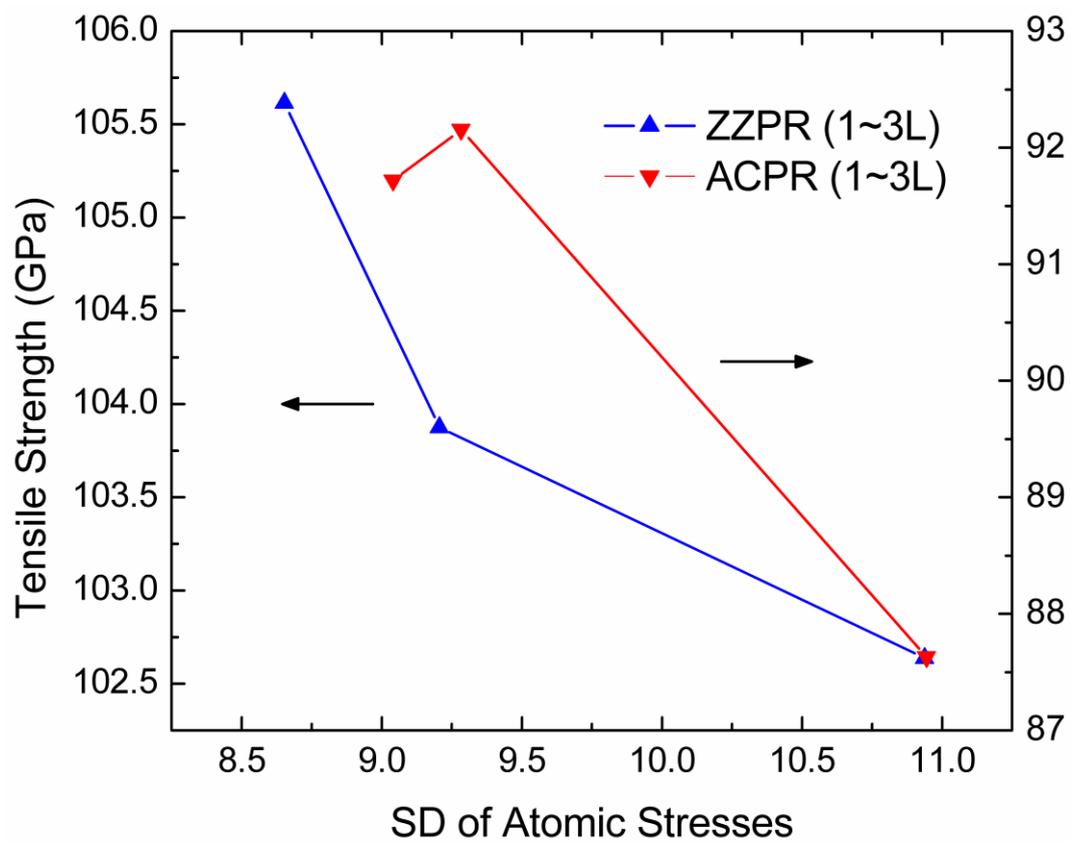

**Figure 5.** The relationship between SD of atomic stress and tensile strength for ZZ-$n$L and AC-$n$L systems ($n=1$~3).

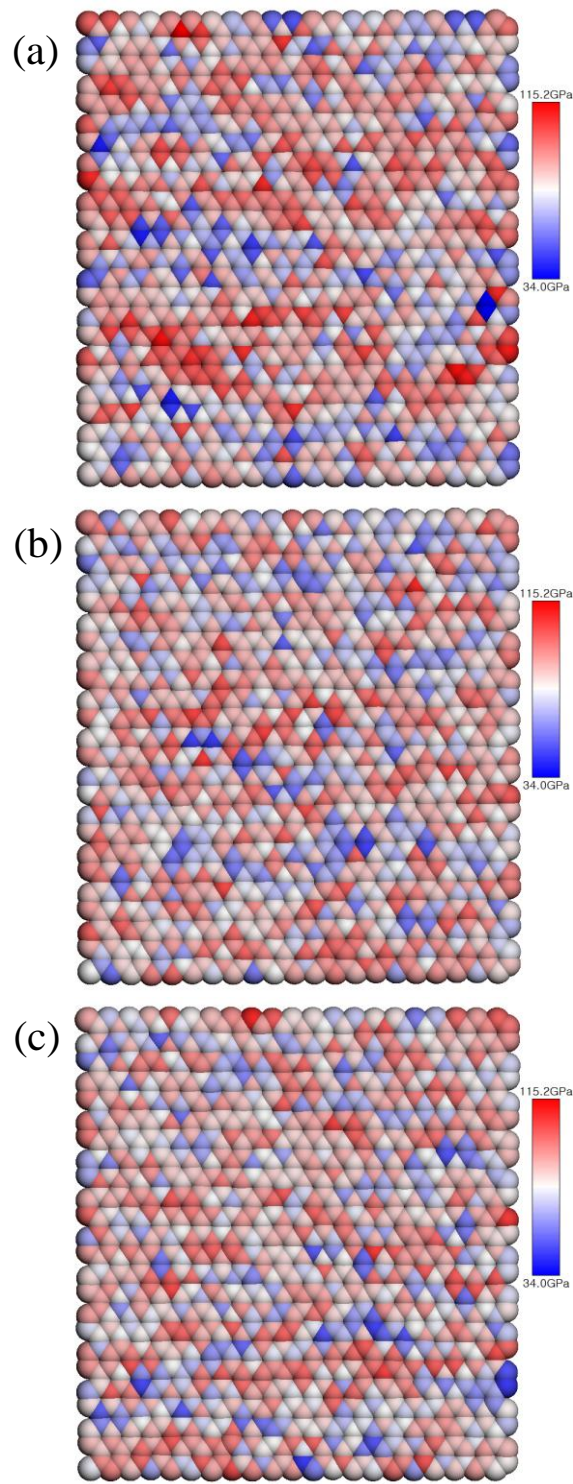

**Figure 6.** The spatial atomic stress distributions of (a) AC-1L, (b) an upper layer of AC-2L, and (c) a lower layer of AC-2L.